\documentclass[preprint,showpacs,preprintnumbers,amsmath,amssymb,nofootinbib]{revtex4}
\usepackage{etex}
\usepackage{graphicx}% Include figure files
\usepackage{dcolumn}% Align table columns on decimal point
\usepackage{bm}% bold math
\usepackage{amssymb,amsthm,amscd,amsbsy,array}
\usepackage{amsmath,dsfont}
\usepackage{soul} 
\usepackage{graphics,xcolor}

\numberwithin{equation}{section}

\newcommand{\half}{{\frac{1}{2}}}
\def\2{{\frac{1}{2}}}
\newcommand{\const}{\mathop{\rm const}\nolimits}
\def\bA{{\bm{A}}}

\def\p{{\partial}}

\def\bR{{\mathds{R}}}

\def\bnabla{\mbox{\boldmath$\nabla$}}
\def\br{{\bm{r}}}

\def\bE{{\bm{E}}}
\def\bB{{\bm{B}}}
\def\bC{{\bm{C}}}

\def\bnabla{{\bm{\nabla}}}

\def\bx{{\bm{x}}}

\def\beq{\begin{equation}}
\def\eeq{\end{equation}}
\def\beqa{\begin{eqnarray}}
\def\eeqa{\end{eqnarray}}
\def\nn{\nonumber}
\def\barray{\left(\begin{array}}
\def\earray{\end{array}\right)}
\def\barraynb{\begin{array}}
\def\earraynb{\end{array}}
\def\SO{{\rm SO}}

\def\smallover#1/#2{\hbox{$\textstyle\frac{#1}{#2}$}} %

\newcommand{\cJ}{\mathcal{J}}

\newcommand{\bbR}{\mathbb{R}}

\newcommand{\fg}{{\mathfrak{{g}}}}

\newcommand{\cV}{{\mathcal V}}
\newcommand{\cM}{{\mathcal M}}

\newcommand{\SL}{\mathrm{SL}}

\newcommand{\hF}{\hat{F}}
\newcommand{\hA}{\widehat{A}}
\newcommand{\hC}{\widehat{C}}
\newcommand{\bone}{{\bf 1}}

\usepackage{color}
\usepackage[colorlinks=true, pdfstartview=FitV, linkcolor=blue, citecolor=blue, urlcolor=blue]{hyperref}

\newcommand{\gb}{\colorbox{green}}

\newcommand{\medbox}[1]{\fbox{%
\rule[-10pt]{0pt}{25pt}$\;\;\displaystyle{#1}\;\;$}%
}
\def\?{{\;\gb{\;\large ?\;}\;}}

\begin{document}

\preprint{%arXiv:
%\red{\bf HeliDual-symplectic-V1} 
}

\title{Duality and helicity: a symplectic viewpoint
}

\author{
M. Elbistan$^{1}$\footnote{mailto:elbistan@impcas.ac.cn.},
C. Duval$^{2}$\footnote{mailto:duval@cpt.univ-mrs.fr},
P. A. Horv\'athy$^{1,3}$\footnote{mailto:horvathy@lmpt.univ-tours.fr},
 P.-M. Zhang$^{1}$\footnote{e-mail:zhpm@impcas.ac.cn},
}

\affiliation{
${}^1$ Institute of Modern Physics, Chinese Academy of Sciences, Lanzhou, (China) 
\\
${}^2$
Aix Marseille Univ, Universit\'e de Toulon, CNRS, CPT, Marseille, France
%Canonical affiliation below:
%Aix-Marseille Universit\'e, CNRS, CPT, UMR 7332, 13288 Marseille, France.\\
%Universit\'e de Toulon, CNRS, CPT, UMR 7332, 83957 La Garde, France.
\\
${}^3$ Laboratoire de Math\'ematiques et de Physique
Th\'eorique,
Universit\'e de Tours,  (France)
}

\date{\today}

\begin{abstract}
The theorem which says that 
helicity is the conserved quantity associated with the duality symmetry of the vacuum Maxwell equations is proved by viewing electromagnetism as an infinite dimensional symplectic system. In fact, it is shown that helicity is the moment map of duality acting as an $\SO(2)$ group of canonical transformations on the symplectic space of all solutions of the vacuum Maxwell equations.
%\\
% \red{\bf HeliDual-symplectic-V1}\\
\end{abstract}

\pacs{\\
11.30.-j 	Symmetry and conservation laws\\
11.30.Cp 	Lorentz and Poincar\'e invariance\\
}

\maketitle

%\tableofcontents
%\newpage

%%%%%%%%%%%%%%%%%%%%%%%%%%%%%%%%%%%%%%%%%%%%%%%%%%%%%%%%%%%%%%%%%%%%%%%%%%%%%%
%%%%%%%%%%%%%%%%%%%%%%%%%%%%%%%%%%%%%%%%%%%%%%%%%%%%%%%%%%%%%%%%%%%%%%%%%%%%%%
\section{Introduction}
%%%%%%%%%%%%%%%%%%%%%%%%%%%%%%%%%%%%%%%%%%%%%%%%%%%%%%%%%%%%%%%%%%%%%%%%%%%%%%
%%%%%%%%%%%%%%%%%%%%%%%%%%%%%%%%%%%%%%%%%%%%%%%%%%%%%%%%%%%%%%%%%%%%%%%%%%%%%%

The usual electromagnetic action in the vacuum,\footnote{Integration is performed over Minkowski spacetime, $M$, endowed with metric $g=g_{\mu\nu}\,dx^\mu{}dx^\nu$ of signature $(+,-,-,-)$. Let us stress that we will content ourselves with a special relativistic treatment of duality, although our main results spelled out in the next sections clearly hold true (with minor modifications) in a fixed gravitational background.} % $g=-\det(g_{\mu\nu})=1$.
\beq
S= -\frac{1}{4}\int_MF_{\mu\nu}F^{\mu\nu} %\sqrt{g}
\,d^4x\,,
\label{emaction}
\eeq
suffers from  well-known nevertheless inconvenient defects, namely the \emph{non-invariance} of the Lagrange density under various symmetry 
 transformations and the consequent non-symmetric form of its energy-momentum tensor, requiring to resort to various ``improvements'' \cite{Jackson,BBN} \footnote{We refer to, e.g., \cite{JMS74} for a geometric standpoint associated with the principle of general covariance, enabling us to circumvent these difficulties.}. In particular, while the vacuum Maxwell equations are invariant w.r.t. \emph{duality trans\-formations}, 
\beq
F\mapsto \hF =  \cos\theta\, F + \sin\theta\,\star(F),
\label{emdual}
\eeq
for any real $\theta$ (where $F={\half} F_{\mu\nu}dx^\mu\wedge dx^\nu$ and 
${\star(F)}=\frac{1}{4}\epsilon_{\mu\nu\rho\sigma} F^{\rho\sigma}dx^\mu\wedge dx^\nu$ is the Hodge dual electromagnetic field strength), the 
 Lagrange density in (\ref{emaction})
is \emph{not invariant}.
The apparent contradiction can be resolved by observing that  a duality rotation
 (\ref{emdual}) changes the Lagrange density by a mere surface term. It is therefore a symmetry of the action \cite{BBN,DeTe} and generates therefore, according to the Noether theorem, a conserved quantity identified here as the optical \emph{helicity} \cite{Calkin}. 
 The proof given in \cite{Calkin} is rather laborious, though, due to the complicated behavior of the vector potential  and the subsequent use of the Hertz vector --- a rather subtle, non-gauge-invariant tool. The treatment  in \cite{DeTe} is also quite involved.
 
Another proposition \cite{Ranada,BBN,Camer3} is to embed the Maxwell theory into a manifestly duality-symmetric one 
for which Noether's theorem yields a seemingly different  expression, namely,\beq
\chi_{_\mathrm{CS}}=\frac{1}{2}\int_{\bR^3}\!(\bA\cdot \bB - \bC\cdot\bE)\,d^3\br
\label{CShel}
\eeq
\textit{\`a la Chern-Simons}, where $\bA$ and $\bC$ are vector potentials for the  magnetic and the electric fields, $\bnabla\times\bA=\bB$ and $\bnabla\times\bC=-\bE$, respectively.
It is worth noting that the second term in Eq. \#~(14) of  \cite{Calkin} and, respectively, in Eq. \#~(2.9) of \cite{DeTe}, both represent the vector potential for the dual field strength --- a fact not recognized by  none of these authors.
 See \cite{AfSt,BBN,Camer3} for comprehensive presentations.
 
\goodbreak

In the first term in (\ref{CShel}) we recognize the 
(magnetic) \emph{helicity}, $\chi_\mathrm{mag}=\frac{1}{2}\int\!\bA\cdot \bB\,d^3\br
 $ widely studied in (magneto)\-hydrodynamics \cite{Moffatt}, where it
 measures the winding of magnetic lines of force and/or fluid vortex lines, respectively.  
It is worth stressing that the magnetic helicity alone is {not} a constant of the motion in general, and the clue leading to (\ref{CShel}) is that its non-conservation,
\beq
\frac{\,d}{dt}\chi_\mathrm{mag}= -\int_{\bbR^3}{
 \bE\cdot\bB \,d^3\br,
 }
\label{FstarF}
\eeq
 is precisely compensated by that of the second term \cite{AfSt}.
A remarkable fact is that (\ref{CShel}) combines two  \emph{Chern-Simons invariants} \cite{ChernSimons}, 
 for both the electromagnetic and its dual field.

Duality and helicity have attracted considerable recent attention, namely in optics \cite{BBN,FeCo,Camer3} and in heavy ion physics \cite{Manuel}. Our own interest stems from  studying the helicity of semiclassical chiral particles \cite{EDHZ-heli}.

In this Note we explain the duality and helicity
from yet another viewpoint, which bypasses Lagrangians and gauge fixing altogether. Our clue is to view the set of solutions of electromagnetism as (an infinite-dimensional)  \emph{symplectic} space \cite{SSD,CW,ABR}. 

%%%%%%%%%%%%%%%%%%%%%%%%%%%%%%%%%%%%%%%%%%%%%%%%%%%%%%%%%%%%%%%%%%%%%%%%%%%%%%
%%%%%%%%%%%%%%%%%%%%%%%%%%%%%%%%%%%%%%%%%%%%%%%%%%%%%%%%%%%%%%%%%%%%%%%%%%%%%%
\section{Electromagnetism in the symplectic framework}
%%%%%%%%%%%%%%%%%%%%%%%%%%%%%%%%%%%%%%%%%%%%%%%%%%%%%%%%%%%%%%%%%%%%%%%%%%%%%%
%%%%%%%%%%%%%%%%%%%%%%%%%%%%%%%%%%%%%%%%%%%%%%%%%%%%%%%%%%%%%%%%%%%%%%%%%%%%%%
 
In the framework of Hamiltonian mechanics \cite{SSD} one works with manifolds  endowed with a closed two-form~$\omega$. If $\dim\ker(\omega)$  has constant but nonzero dimension, $\omega$ is called presymplectic; if its kernel is zero dimensional, it is called symplectic. In the physical applications we have in mind, we start with a manifold  such that
$(\cV,\omega)$ is presymplectic and is referred to as an \textit{``evolution space''}, where the dynamics takes place. The characteristic leaves which integrate $\ker(\omega)$ are identified with the motions of the system.
The quotient of $\cV$ by the characteristic foliation 
of $\omega$, namely $\cM=\cV/\ker(\omega)$, is therefore endowed with a symplectic two-form $\Omega$, whose pull-back to $\cV$ is $\omega$. Then $(\cM,\Omega)$
is what has been called the \textit{``space of motions''} in \cite{SSD}. 
Crnkovi\v{c} and Witten \cite{CW} call it  the ``true phase space''.

The next ingredient is a Lie group $G$ of canonical transformations, i.e., of diffeomorphisms of $\cV$ preserving the two-form~$\omega$. Denote by $\fg$ the Lie algebra of $G$, and by $Z_{\cV}$ the infinitesimal action (fundamental vector field) on $\cV$ associated with $Z\in\fg$. 

We thus have $L_{Z_{\cV}}\omega=0$ so that $\omega(Z_{\cV},\,\cdot\,)$ is a closed one-form for all $Z\in\fg$.
We now say that $J:\cV\to\fg^*$ is a \emph{moment map} for $(\cV,\omega,G)$ if the  stronger condition  
\begin{equation}\label{J}
\omega(Z_{\cV},\,\cdot\,)=-d(J\cdot{}Z)
\end{equation} 
holds for all $Z\in\fg$.\footnote{For each point $x$ of $V$, the quantity $J(x)$ belongs to the dual $\fg^*$ of the Lie algebra $\fg$, and contracting with $Z\in\fg$ yields a function $x\mapsto{}J(x)\cdot{}Z$ on $\cV$.}

If the equations of motion are given by $\ker(\omega)$, as it happens in the mechanics of finite dimensional systems \cite{SSD} and, as we will prove below, also for Maxwell's electromagnetism, then~$J$ clearly descends to the space of motions, $\cM=\cV/\ker(\omega)$, as the \emph{Noetherian quantity}  as\-sociated with the symmetry group $G$~: indeed
  $J\cdot{}Z$ is a \emph{constant of the motion} for all~$Z\in\fg$.

Below we boldly extend this framework  to the infinite dimensional  ``manifold'' $\cM$ which consists of all \emph{solutions} of the vacuum Maxwell equations modulo gauge transformations we endow with a \emph{symplectic structure}.\footnote{A rigorous treatment of this infinite-dimensional differentiable structure would  require the use of, e.g., diffeology \cite{PIZ}, especially when dealing with differential forms on this ``diffeological space''.}

Let us show how all this comes about. Our first aim is to translate the usual variational approach into a symplectic language.
 The actual physical variable is the potential one-form $A=A_\mu\,dx^\mu$ locally defined by $F=dA$.\footnote{One-forms and vector fields are identified by lifting and lowering indices using the Minkowski metric.} Then the  variation of the action (\ref{emaction}) with respect to a variation  $\delta A=\delta A_{\mu}\,dx^\mu$ of the $4$-potential is
\beq
\delta S =
\int_M\big[\p_\nu (F^{\mu\nu}\delta A_\mu)+(\p_\mu F^{\mu\nu})\delta A_\nu\big] 
\,d^4x\,.
\label{Maxvar}
\eeq
Assuming that the fields drop off sufficiently rapidly at infinity  --- or that the variations~$\delta A$ have compact support --- the surface term can be dropped, allowing us to deduce the vacuum Maxwell equations 
$
\p_{[\mu}F_{\nu\rho]}=~0
$
and
$\p_\mu F^{\mu\nu}=0$, 
also  written as
\beq
dF=0
\qquad
\hbox{and}
\qquad
d\star(F)=0.
\label{Maxeqn}
\eeq
%%%%%%%%%
Denote by ${\cV}$ the space of one-forms $A$ of Minkowski space $M$ whose associated field strength, $F=dA$, is a \emph{solution} of (\ref{Maxeqn}). We contend that ${\cV}$, which can be thought of as an
infinite-dimensional manifold (affine space), is an ``evolution space" for the Maxwell theory.

Firstly, a variation of a \emph{solution}, $\delta A$, is a ``tangent vector'' to ${\cV}$ at $A\in{\cV}$ if
$A+\delta A$ is still a solution of the field equations  which vanishes at spatial infinity (as $A$ does). Since the associated field strength is
%$
%%\p_{\mu}(A_{\nu}+\delta A_\nu) -
%%\p_{\nu}(A_{\mu}+\delta A_\mu)=
%F_{\mu\nu}+\delta F_{\mu\nu}
%$
$F+\delta F$, 
%$\delta F_{\mu\nu} = \p_{\mu}\delta A_\nu -
%\p_{\nu}\delta A_\mu$,
where $\delta F = d(\delta A)$,
 it follows that $\delta F$
also satisfies the Maxwell equations,
$ 
d(\delta F)=0$ and $d\star(\delta{F})=0.
$ 

Now, adapting Souriau's procedure in \cite{SSD}, Sec. 7, to field theory, we define a symplectic form on the space of all solutions of the linear system (\ref{Maxeqn}). To this end,  
we consider the action (\ref{emaction}) by integrating over the domain $M'=[t_0,t_1]\times\Sigma\subset{M}$ defined by a Cauchy $3$-surface $\Sigma$ with \emph{arbitrary dates}~$t_0$ and $t_1\neq{}t_0$, where $t$ is some given time-function. When~$F$ is a solution of the Maxwell equations, the variation  vanishes, $\delta S=0$, and therefore Eq.~(\ref{Maxvar}) boils down to
\beq
0=
\int_M\p_\nu (F^{\mu\nu}\delta A_\mu)\,d^4x=
%\int_{\Sigma_1}F^{\mu\nu}\delta A_\mu\,d^3\br-
%\int_{\Sigma_0}F^{\mu\nu}\delta A_\mu\,d^3\br
%\int_{\Sigma_1}{\alpha(\delta A)}-\int_{\Sigma_0}{\alpha(\delta A)}
\int_{\Sigma_1}\!\star(F(\delta{A}))
-
\int_{\Sigma_0}\!\star(F(\delta{A}))
\,,
\nn
\eeq
where $\Sigma_i=\{t_i\}\times\Sigma$ for $i=0,1$, implying that the integral does not depend on the choice of $t_0$ and $t_1$; the one-form\footnote{
In a coordinate system where the metric is $g=dt^2-d\bx^2$ and $\Sigma$ given by $t=\const$, Eq. (\ref{CartanMax}) reads
\begin{equation}
\label{CartanMaxBis}
\alpha(\delta{A})=
\int{\! 
F^{\mu\nu}\delta A_\mu\p_\nu t\,d^3\bx}.
\end{equation}
}
\begin{equation}
\label{CartanMax}
\medbox{
\alpha(\delta{A})=
\int_\Sigma{\!\star(F(\delta{A}))}
=
-\int_\Sigma{\star(F)\wedge\delta A}
}
\end{equation}
 is therefore well-defined;  it is the \emph{Cartan one-form}. 
 The expression~(\ref{CartanMax}) represents the \emph{flux} of the vector field $F(\delta A)=(F^{\mu\nu}\delta A_\mu)\p_\nu$ across the Cauchy surface $\Sigma$.  
%%%%
Calculating the exterior derivative, $\omega=d\alpha$,  via $d\alpha(\delta A,\delta'\!{A})=\delta(\alpha(\delta'\!{A}))-\delta'(\alpha(\delta A))-\alpha([\delta,\delta']A)$,
we find 
\begin{equation}
\label{2FormMaxBis}
\omega(\delta{A},\delta'\!{A})=
\int_\Sigma\! \delta{A}\wedge\star(\delta'F)-\delta'\!{A}\wedge\star(\delta F).
\end{equation}
 The two-form (\ref{2FormMaxBis}) corresponds \emph{exactly} to that given by Eq. \# (23) in~\cite{CW}.
 
From this point on, we do not use any Lagrangian; the starting point of all our subsequent investigations will be the two-form (\ref{2FormMaxBis}).

Let us now show that $({\cV},\omega)$ becomes a formal \emph{presymplectic space}. To that end, let us compute its characteristic distribution.
We thus must determine the kernel of $\omega$, i.e., all variations $\delta A$ of a solution $A\in{\cV}$ such that $\omega(\delta A,\delta'\!{A})=0$ for all $\delta'\!{A}$, subject to the constraint $\delta'(d\star(F))=0$ to comply with the field equations. Using a Lagrange multiplier,~$f$, we look for all solutions $\delta A$~of
\begin{eqnarray}
\int_\Sigma{\! \delta{A}\wedge\star(\delta'F)-\delta'\!{A}\wedge\star(\delta F)}
%\nonumber
%&=&
=
-\int_\Sigma{\!f\,d(\star(\delta'F))}
%\\
%&=&
=
\int_\Sigma{\!df\wedge{}\star(\delta'F)}
\label{Eq2}
\end{eqnarray}
for all compactly supported variations $\delta'\!{A}$. 
Eq. (\ref{Eq2}) readily yields that the kernel is indeed given by all gauge transformations,
\begin{equation}\label{keromega}
\medbox{
\delta A\in\ker(\omega) 
\quad
\iff
\quad
\delta A=df
}
\end{equation}
for some smooth function $f$. (Note that we duly have $\delta F=0$.)
Then, the leaves of the characteristic distribution $\ker(\omega)$ are identified to the orbits of the \emph{electromagnetic gauge group} $\cJ$ generated by smooth functions $\varphi$ of $M$, which acts on ${\cV}$ according to
$A\mapsto {}A+d\varphi$.
At last, the quotient
\begin{equation}\label{MaxSympl}
\medbox{
\cM={\cV}/\cJ
}
\end{equation}
is the \emph{the \emph{``space of motions''} of electromagnetism; it is identified with the space  of all vector potentials which are solutions of the free Maxwell equations modulo gauge transformations},
to which $\omega$ projects as the canonical \emph{symplectic two-form} $\Omega$.

%%%%%%%%%%%%%%%%%%%%%%%%%%%%%%%%%%%%%%%%%%%%%%%%%%%%%%%%%%%%%%%%%%%%%%%%%%%%%%
%%%%%%%%%%%%%%%%%%%%%%%%%%%%%%%%%%%%%%%%%%%%%%%%%%%%%%%%%%%%%%%%%%%%%%%%%%%%%%
\section{Duality symmetry}
%%%%%%%%%%%%%%%%%%%%%%%%%%%%%%%%%%%%%%%%%%%%%%%%%%%%%%%%%%%%%%%%%%%%%%%%%%%%%%
%%%%%%%%%%%%%%%%%%%%%%%%%%%%%%%%%%%%%%%%%%%%%%%%%%%%%%%%%%%%%%%%%%%%%%%%%%%%%%

%%%%%%%%%%%%%%%%%%%%%%%%%%%%%%%%%%%%%%%%%%%%%%%%%%%%%%%%%%%%%%%%%%%%%%%%%%%%%%
%\subsection{Duality as a canonical transformation}
%%%%%%%%%%%%%%%%%%%%%%%%%%%%%%%%%%%%%%%%%%%%%%%%%%%%%%%%%%%%%%%%%%%%%%%%%%%%%%

Let us now consider duality rotations (\ref{emdual}) which form, as said before, a manifest symmetry group for the free Maxwell equations.\footnote{
%To trace the origin of this group, 
%Let us recall that
The field equations being linear, any real linear transformation $\hF=a F + b \star(F)$ \& $\star(\hF)= c F+d\star(F)$, with $ad-bc\neq0$, permutes the solutions of (\ref{Maxeqn}). Now, the Hodge star defines a complex structure on the $2$-dimensional space spanned by $F$ and $\star(F)$, since
$\star^2=-\bone$.
Restricting our considerations to transformations that preserve the ``star'' $\star$, i.e., to $\mathrm{Sp}(1,\bbR)\cong\SL(2,\bbR)$, an easy calculation 
shows  that $c=-b$ and $d=a$, implying $a^2+b^2=1$; hence $a=\cos\theta$ and $b=\sin\theta$ as in Eq. (\ref{emdual}). }
%%%%%%%%%%
Using our symplectic language,
we claim that the two-form $\omega$ in (\ref{2FormMaxBis}) is invariant under the  (\ref{emdual}), implemented on the potentials as
\begin{equation}
\label{hA}
\hA=\cos\theta\, A + \sin\theta\, C,
\qquad
\hC=\cos\theta\, C - \sin\theta\, A
\end{equation}
where $A$ and $C$ are (local) $4$-potentials for the field and its dual, $F=dA$ and $\star(F)=dC$. Note that $A$ and $C$ here are \emph{not} independent since their field strengths are each other's duals.
Using the properties of the Hodge star operation, $\star$, one shows indeed that
\begin{equation}\label{Symplectomorphism}
\medbox{
\omega(\delta{\hA},\delta'\hA)
=
\omega(\delta{A},\delta'A)\,
}
\end{equation}
for all variations $\delta A$ and $\delta'A$ compatible with the constraints (\ref{Maxeqn}). 
This proves that
\emph{the duality transformation (\ref{emdual}), implemented as above 
is a canonical trans\-formation of the evolution space, 
$({\cV},\omega)$, and therefore also of the space of motions, $(\cM,\Omega)$}.

%%%%%%%%%%%%%%%%%%%%%%%%%%%%%%%%%%%%%%%%%%%%%%%%%%%%%%%%%%%%%%%%%%%%%%%%%%%%%%
%\subsection{The duality moment map}
%%%%%%%%%%%%%%%%%%%%%%%%%%%%%%%%%%%%%%%%%%%%%%%%%%%%%%%%%%%%%%%%%%%%%%%%%%%%%%

We now turn to the \emph{moment map} of duality symmetry. The infinitesimal duality action on ${\cV}$ is given by
$ 
\delta_\varepsilon{A}=\varepsilon\, C
$ and $
\delta_\varepsilon{C}=-\varepsilon\, A,
$ 
where $\varepsilon\in\bbR$. 
A straightforward calculation then shows that, for all $\delta'A$ compatible with the constraints (\ref{Maxeqn}), we have
\begin{eqnarray}
\omega(\delta_\varepsilon{A},\delta'\!A)
=
\int_\Sigma{\left\{
\delta'(\star(F))\wedge\varepsilon{}C+\varepsilon{}F\wedge\delta'\!A
\right\}}
=
\half\varepsilon\,\delta'\!\!\int_\Sigma{\left\{
C\wedge\star(F)
+
A\wedge{}F
\right\}}
\end{eqnarray}
since $\delta'\!A\wedge{}F\equiv\half\delta'(A\wedge{}F)$ and, likewise, 
$\delta'C\wedge\star(F)\equiv\half\delta'(C\wedge\star(F))$ \--- modulo an exact three-form.
It follows that we \emph{do actually have a moment map}
$J:{\cV}\to\bbR$, i.e., such that
$ 
\omega(\delta_\varepsilon{A},\delta'A)=-\delta'\big(J(A)\varepsilon\big)
$ 
for the  duality group acting on $(\cV,\omega)$, and thus on the space of motions of all solutions of the Maxwell equations,
namely
\begin{equation}
\label{MaxMomentMap}
\medbox{
J(A) = -\half\int_\Sigma{A\wedge{}dA+C\wedge{}dC\,,\,}
}
\end{equation}
which is indeed the geometric form of
% ``Chern-Simons'' 
 of the helicity,  (\ref{CShel}). The conservation of (\ref{MaxMomentMap}) can also be checked directly: the two Chern-Simons three-forms are both the  anti-derivatives of the \emph{same} Pontriagin density, but with \emph{opposite signs},
\beq
d\big(A\wedge F\big) = F\wedge{}F = -\star(F)\wedge\star(F) = 
-d\big(C\wedge \star(F)\big).
\label{Pontriagin}
\eeq
Let us consider two Cauchy surfaces $\Sigma_0$ and $\Sigma_1$ with dates $t=t_0$ and $t=t_1$ and view them as the boundaries of a four-volume $V$. The integral of the four-form $-\half{}d\big(A\wedge F+C\wedge\star(F)\big)$ on~$V$ vanishes in view of (\ref{Pontriagin}), proving that the fluxes across $\Sigma_0$ and $\Sigma_1$ are equal, and that the moment map $J$ in (\ref{MaxMomentMap}) is therefore independent of $\Sigma$.

%\begin{redtext}
The equivalence of (\ref{MaxMomentMap}) with the optical formula in the literature 
which says that \emph{the optical helicity is in fact the difference of the left- and right-handed photons},
\beq
\chi_{\mathrm{O}}=N_L-N_R,
\label{optihel}
\eeq
 can be shown along the lines followed in \cite{Calkin,AfSt}.
 
Here we just mention an alternative yet incomplete approach~: the general form (\ref{CShel}) was narrowly missed by Ra\~nada \cite{Ranada}, who did correctly identify both terms --- without adding them however, and  considering  only the special case $\bE\cdot\bB=0$, when both terms are separately conserved.
 cf. (\ref{FstarF}). Under such  condition he could show that the two integrals are indeed the degrees, $N_L$ and $N_R$, of suitable Hopf maps $S^3\to{}S^2$, confirming (\ref{optihel}) in such a case. Extension of this approach to the general case is under investigation.

%%%%%%%%%%%%%%%%%%%%%%%%%%%%%%%%%%%%%%%%%%%%%%%%%%%%%%%%%%%%%%%%%%%%%%%%%%%%%%
%%%%%%%%%%%%%%%%%%%%%%%%%%%%%%%%%%%%%%%%%%%%%%%%%%%%%%%%%%%%%%%%%%%%%%%%%%%%%%
\section{Conclusion}
%%%%%%%%%%%%%%%%%%%%%%%%%%%%%%%%%%%%%%%%%%%%%%%%%%%%%%%%%%%%%%%%%%%%%%%%%%%%%%
%%%%%%%%%%%%%%%%%%%%%%%%%%%%%%%%%%%%%%%%%%%%%%%%%%%%%%%%%%%%%%%%%%%%%%%%%%%%%%

In this ``variation on a themes''-type Note
 we re-derive, using the symplectic framework in infinite dimensions, the helicity formula (\ref{MaxMomentMap}), equivalent to the one 
  (\ref{CShel}) proposed in the literature. Unlike for previous authors \cite{BBN, Calkin, AfSt}, our derivation is gauge-invariant, as it did not require any choice of gauge.

We note also that our two-form (\ref{2FormMaxBis}) is manifestly duality-invariant, whereas the Cartan one-form $\alpha$ in~(\ref{CartanMax}) is clearly \emph{not}, as it follows from the non-invariance of the standard Maxwell Lagrangian (\ref{emaction}).  
This   highlights the advantage of using the presymplectic Maxwell two-form~(\ref{2FormMaxBis}) to deal with symmetries, and in particular with duality.

The situation is reminiscent of what happens for a Dirac monopole, for which  no manifestly radially symmetric vector potential and thus no symmetric Lagrangian or Cartan one-form can exist, whereas the two-form which represents the field strength resp. the dynamics is perfectly rotationally invariant \cite{HPA81}.

We would also mention that this formula can also be obtained using the Pauli-Lubanski  approach \cite{BB-Perjes}, also followed in \cite{EDHZ-heli}.

%%%%%%%%%%%%%%%%%%%%%%%
\begin{acknowledgments} 
CD warmly thanks  H. P. K\"unzle and M. J. Gotay for enlightening discussions at the early stage of this work. PH would like to thank J. Balog and K. Bliokh for discussions. ME and PH are grateful to the IMP of the CAS for hospitality in Lanzhou. 
This work was supported by the Major State Basic Research Development Program in China (No.
2015CB856903), the National Natural Science Foundation of China (Grant No. 11575254 and 11175215).
\end{acknowledgments}

\bigskip\goodbreak\newpage
%\bigskip\goodbreak\newpage

%%%%%%%%%%%%%%%%%%%%%%%%%%%%%%%%%%%%%%%%%%%%%%%%%%%%%%%%%%%%%%%%%%%%%%%%%%%%%%
%%%%%%%%%%%%%%%%%%%%%%%%%%%%%%%%%%%%%%%%%%%%%%%%%%%%%%%%%%%%%%%%%%%%%%%%%%%%%%

\end{document}